\documentclass[12pt,preprint]{aastex}
\usepackage[dvips]{epsfig}
\usepackage[dvips]{color}

\begin{document}
\shorttitle{\sc AN EXTREMELY CARBON-RICH METAL-POOR STAR IN SEGUE 1}
\shortauthors{NORRIS ET Al.}

\newcommand{\seg}     {Segue~1}
\newcommand{\segs}    {Seg~1--7}
\newcommand{\cs}      {CS22957--027}
\newcommand{\boo}     {Bo\"{o}tes~I}
\newcommand{\boos}    {Boo$-$1137}
\newcommand{\kms}     {\rm km~s$^{-1}$}
\newcommand{\teff}    {$T_{\rm eff}$} 
\newcommand{\logg}    {log~$g$} 
\newcommand{\loggf}   {log~$gf$} 
\newcommand{\msun}    {M$_{\odot}$}
\newcommand{\lsun}    {L$_{\odot}$}
\newcommand{\alfe}    {[$\alpha$/Fe]} 
\newcommand{\uf}      {ultra-faint}
\newcommand{\rh}      {r$_{h}$}

\title{AN EXTREMELY CARBON-RICH, EXTREMELY METAL-POOR STAR IN THE SEGUE
1 SYSTEM\footnote{Observations obtained for ESO program P383.B-0038,
using VLT-UT2/UVES}}

\author {JOHN E. NORRIS\altaffilmark{1}, GERARD
GILMORE\altaffilmark{2}, ROSEMARY F.G. WYSE\altaffilmark{3}, DAVID
YONG\altaffilmark{1}, AND ANNA FREBEL\altaffilmark{4} }

\altaffiltext{1}{Research School of Astronomy \& Astrophysics, The
Australian National University, Mount Stromlo Observatory, Cotter
Road, Weston, ACT 2611, Australia; email: jen@mso.anu.edu.au}

\altaffiltext{2}{Institute of Astronomy, University of Cambridge,
Madingley Road, Cambridge CB3 0HA, UK}

\altaffiltext{3}{The Johns Hopkins University, Department of Physics
\& Astronomy, 3900 N.~Charles Street, Baltimore, MD 21218, USA}

\altaffiltext{4}{Harvard-Smithsonian Center for Astrophysics,
Cambridge, MA 02138, USA}

\begin{abstract}

We report the analysis of high-resolution, high-$S/N$ spectra of an
extremely metal-poor, extremely C-rich red giant, {\segs}, in the
{\seg} system -- described in the literature alternatively as an
unusually extended globular cluster or an {\uf} dwarf galaxy.  The
radial velocity of {\segs} coincides precisely with the systemic
velocity of {\seg}, and its chemical abundance signature of [Fe/H] =
--3.52, [C/Fe] = +2.3, [N/Fe] = +0.8, [Na/Fe] = +0.53, [Mg/Fe] =
+0.94, [Al/Fe] = +0.23 and [Ba/Fe] $<$ --1.0 is similar to that of the
rare and enigmatic class of Galactic halo objects designated CEMP-no
(Carbon-rich, Extremely Metal-Poor and with no enhancement (over solar
ratios) of heavy neutron-capture elements).  This is the first star
in a Milky Way ``satellite'' that unambiguously lies on the
metal-poor, C-rich branch of the Aoki et al.\ (2007) bimodal
distribution defined by field halo stars in the ([C/Fe],
[Fe/H])--plane.  Available data permit us only to identify {\segs} as
a member of an {\uf} dwarf galaxy or as debris from the Sgr dwarf
spheroidal galaxy.  In either case, this demonstrates that at
extremely low abundance, [Fe/H ] $ < -3.0$, star formation and
associated chemical evolution proceeded similarly in the progenitors
of both the field halo and satellite systems. By extension, this is
consistent with other recent suggestions the most metal-poor dwarf
spheroidal and {\uf} dwarf satellites were the building blocks of the
Milky Way's outer halo.

\end{abstract}


\section{INTRODUCTION}

Studies of the Milky Way's dwarf spheroidal and {\uf} satellite
galaxies are placing important constraints on the role these systems
have played in the formation of its halo populations.  Early work on
stars in the classical dwarf spheroidal galaxies (dSph), concerning
the abundance of the $\alpha$ elements Mg and Ca (relative to Fe) in
the range --2.0 $<$ [Fe/H] $<$ --1.0, established that their {\alfe}
values are significantly lower than in the bulk of Galactic halo stars
in the solar neighborhood at the same [Fe/H] -- suggesting that there
is a fundamental difference between the halo and its dwarf galaxies
(see e.g. Tolstoy, Hill \& Tosi 2009, and references therein). These
lower values of {\alfe} reflect the extended star-formation histories
of the host systems (see Gilmore \& Wyse 1991; Unavane, Wyse \&
Gilmore 1996). Recently, however, work by Nissen \& Schuster (2010) on
field halo stars, testing the suggestion of Carollo et al.\ (2007,
2010) that the Galaxy's halo comprises inner and outer components
having distinct chemical abundance distributions and kinematics,
clearly established there are two distinct components in the ({\alfe},
[Fe/H])-plane at --1.6 $<$ [Fe/H] $<$ --0.8, in the sense that stars
in retrograde Galactic orbits (which Carollo et al. identify as
characteristic of the outer halo) have lower {\alfe} than those on
prograde orbits (predominantly the inner halo).  Another relevant
result is that at [Fe/H] $\sim$ --3.7 in these dwarf galaxies the
relative abundances of a large number of elements are quite similar to
those found in the majority of Galactic halo red giants (Frebel,
Kirby, \& Simon 2010a (Sculptor: 8 elements); Norris et al. 2010a
({\boo}: 15 elements)).

Parallel investigations on the relative abundances of carbon ([C/Fe])
below [Fe/H] = --3.0 by Frebel et al. (2010b) and Norris et al.\
(2008, 2010b) demonstrate a large range in [C/Fe] in the {\uf} dwarf
galaxies, with similar [C/Fe] patterns in the {\uf} systems and in the
Galactic halo.  The purpose of the present Letter is to investigate
further the result of Norris et al.  (2010b), based on intermediate
resolution spectroscopy, that in the extreme {\uf} {\seg} system
(M$_{V,\,total}$ $\sim$~--1.5, baryonic mass $\sim$~1000{\msun};
Belokurov et al.\ 2007, Martin et al.\ 2008) there is an extremely
metal-poor giant, {\segs}, ([Fe/H] $\sim$~--3.5) which is also
extremely carbon rich ([C/Fe] $\sim$~+2.3).  Further, {\segs} lies
almost 4 nominal half-light radii from the center of {\seg} and its
membership has important implications for the structure and nature of
the system.

\section{OBSERVATIONS}

{\segs} has coordinates $\alpha (2000) = 10$\,h $08$\,m $14.4$\,s and
$\delta (2000) = +16^{\circ}\;05'\;01''$.  $ugriz$ data are available
from Data Release 7 of the Sloan Digital Sky Survey (Abazajian et al.\
2009).  Adopting E(B--V) = 0.032 (Geha et al.\ 2009) and following
Schlegel, Finkbeiner, \& Davis (1998), we obtain (g--r)$_{0}$ = 0.596
$\pm$ 0.009 and (r--z)$_{0}$ = 0.354 $\pm$ 0.012 for {\segs}.

\subsection{High-resolution Spectroscopy}

{\segs} was observed in Service Mode at the Very Large Telescope (VLT)
Unit Telescope 2 (UT2) with the Ultraviolet-Visual Echelle
Spectrograph (UVES) (Dekker et al.\ 2000)
on five nights during the period 2010 January 27 through March 12,
during which 11 individual exposures were obtained, each having
integration time of 46\,min. The instrumental setup was as described
in our investigation of {\boos}, a red giant member of {\boo} (Norris
et al.\ 2010a, hereafter NYGW), and will not be repeated here.
Suffice it to say that spectra were obtained over the wavelength
ranges 3300--4520\,{\AA}, 4620--5600\,{\AA}, and 5680--6650\,{\AA}.
The 11 pipeline-reduced spectra were co-added to produce the final
spectra which have $S/N$ per $\sim0.03$\,{\AA} pixel increasing from
25 to 50 over 3700--4500\,{\AA}, roughly constant at $\sim$~65 over
4620--5600\,{\AA}, and increasing from 70 to 100 over
5680--6650\,{\AA}.

The spectra of {\segs} have weak metal lines and absorption dominated
by features of the CH molecule, bearing a strong resemblance to the
extremely metal-poor, C-rich, CEMP-no star{\footnote{ CEMP-no stars
are metal-poor objects having [C/Fe] $>$ 1.0 and [Ba/Fe] $<$ 0.0, the
``no'' referring to the lack of enhancement in neutron-capture elements
(Beers \& Christlieb 2005, their Table 2).} CS22957--027, first
analyzed at high resolution and high $S/N$ by Norris, Ryan \& Beers
(1997).  The nature of the spectra is shown in
Figure~\ref{Fig:Spectra}, where in the upper panel {\segs} is compared
with {\cs} and the ``normal'' extremely metal-poor red giant {\boos},
all of which have very similar {\teff}, {\logg}, and [Fe/H] as shown
in the figure.  For heuristic purposes, these spectra have been
broadened with a Gaussian having FWHM = 1.0~{\AA}.  The important
point of the panel is the enormous CH blanketing present in {\segs}
and {\cs}, and its absence from {\boos}.  As colleague S.G. Ryan wrote
upon first observing {\cs} at high resolution in this wavelength
region: ``Actually, it's hard to see anything except CH."  The lower
panel in Figure~\ref{Fig:Spectra} presents the original, unbroadened,
spectra over the wavelength range 3900--4000~{\AA}.  We make three
comments on these spectra.  First, the similarity in the strength of
the broad Ca II H \& K lines at 3933.6~{\AA} and 3968.4~{\AA},
suggests the two stars have similar heavy element abundances.  Second,
most of the other lines are due to CH.  Third, several of the features
appear double in {\cs} but not in {\segs}.  We shall return to this
point in Section 3, but note here that the extra features in {\cs} are
$^{13}$CH lines and result from the lower $^{12}$C/$^{13}$C ratio of
{\cs} compared with that of {\segs}.

\subsection{Line Strength Measurements}

Line strengths for atomic species were measured by J.E.N., as
described in some detail in NYGW.  Results are presented in Table 1,
where columns (1) and (2) give identification, and (3)--(5) contain
lower excitation potential, {\loggf}, and equivalent widths.  The
atomic data are from Cayrel et al.\ (2004), except for one line as
indicated in the table.  There was one essential difference between
the present and earlier work: because the blue part of the {\segs}
spectrum is heavily contaminated by CH lines we followed our earlier
procedure for {\cs} (see Norris et al.\ 1997) and used spectrum
synthesis techniques to determine when one might expect blending of CH
lines with atomic features to contaminate the latter.  We did not
measure atomic features when such contamination was possibly present.

\subsection{Radial Velocity}

We measured the radial velocity for each of our 11 observations using
cross correlation techniques against a model synthetic spectrum as
described by NYGW.  The resulting heliocentric radial velocity is
204.3 $\pm$ 0.1 {\kms} (internal error), while the range in velocities
was 203.7--204.6 {\kms}, suggesting that the velocity of {\segs} was
essentially constant over the six week period covered by the
observations.  The external error in the velocity is $\sim$0.5
{\kms}. According to Simon et al.\ (2010) the systemic velocity of
{\seg} is 208.5 $\pm$ 0.9~{\kms}, with dispersion
3.7$^{+1.4}_{-1.1}$~{\kms}.  Our radial velocity data are thus
consistent with {\segs} being a member of the system.

\section{ABUNDANCE ANALYSIS}

The effective temperature and surface gravity of {\segs}, based on
$ugriz$ photometry, the synthetic $ugriz$ colors of
Castelli\footnote{\texttt{http://wwwuser.oat.ts.astro.it/castelli/colors/sloan.html}},
and the Yale--Yonsei (YY) Isochrones (Demarque et al.\ 2004) are
{\teff} = 4960 $\pm$ 140~K and {\logg} = 1.9 $\pm$ 0.4 (Norris et al.\
2010b).  Adopting procedures described in NYGW, we analyzed the
equivalent widths in Table~1\footnote{Our abundance for Si is based on
Si~I~4102.94\,{\AA} (slightly contaminated by the wing of H$\delta$)
for which we adopted spectrum synthesis techniques.}, together with
the (0,0) and (1,1) bands of the $A-X$ electronic transition of the CH
molecule in the interval 4250--4330\,{\AA}, and the (0,0) and (1,1)
bands of the $A-X$ electronic transition of the NH molecule in the
range 3340--3400\,{\AA} using a 2009 version of the LTE stellar line
analysis program MOOG (Sneden 1973), modified to include a proper
treatment of continuum scattering (Sobeck et al.\ 2010 in prep.),
together with the 1D, LTE model atmosphere atmospheres of Castelli \&
Kurucz (2003) to determine abundances for 14 elements, together with
limits for a further three.  (In our earlier work we established that
these techniques produce abundances for atomic species in excellent
agreement with those of Cayrel et al.\ (2004), when applied to their
data set (see NYGW, Section 3.2)).  For {\segs}, our analysis yielded
a microturbulent velocity $\xi_{\rm t}$ = 1.3 {\kms} and the
abundances presented in Table~2\footnote{While this microturbulence is
somewhat lower than expected, we note that of the 35 extremely
metal-poor red giants analyzed by Cayrel et al\ (2004, their Table 5)
six have $\xi_{\rm t}$~$\leq$~1.5~{\kms}.}, where columns (1)--(5)
contain species, number of lines measured (or, alternatively, that
synthetic spectrum techniques were adopted), log($\epsilon$(X) (=
log(N(X)/N(H))$_{\rm{star}}$ + 12.00), its error, and the relative
abundance [X/Fe], respectively.  The final column in the table
contains the total error in relative abundance, determined by
quadratic addition of the internal error in column (5) and the
systematic errors caused by errors in atmospheric parameters --
$\sigma$({\teff}) = 140~K, $\sigma$({\logg}) = 0.4, $\sigma$([Fe/H]) =
0.3, and $\sigma$($\xi_{\rm t}$) = 0.3 {\kms} -- following NYGW.
Inspection of Table 2 confirms that, with [Fe/H] = --3.52, [C/Fe] =
+2.3 and [Ba/Fe] $<$ --1.0, {\segs} is indeed a CEMP-no star.

Finally, as noted in Section 2.1, the features of $^{13}$CH are not
seen in {\segs}, in contrast to their clear presence in {\cs}, which
has $^{12}$C/$^{13}$C = 10 $\pm$ 5 (Norris et al.\ 1997).  For {\segs}
we estimate a lower limit for $^{12}$C/$^{13}$C of 50 $\pm$ 10 from
the CH features at 4019 and 4237~{\AA}.

\section{DISCUSSION}

\subsection{Relative Abundances of {\segs} and the Galactic halo C-rich stars}

Figure~\ref{Fig:DXFe}} presents the difference in relative abundances,
$\Delta$[X/Fe], between {\segs}, Galactic halo CEMP-no red giants
having --4.0 $<$ [Fe/H] $<$ --3.0, and the C-normal {\boos}, on the
one hand, and the ``unmixed'' Galactic halo red giant branch (RGB)
stars of Spite et al. (2005)\footnote{We adopt the average values of
[X/Fe] for the 12 ``unmixed'' RGB stars in Table 3 of Spite et
al. (2005) that have --4.0 $<$ [Fe/H] $<$ --3.0, excluding the CEMP-r
star CS22892-052. These objects are chosen to minimize the effect of
internal mixing that occurs above the so-called ``RGB bump'' in the
color magnitude diagram of metal-poor red giants (Gratton et al.\
2000; Spite et al.\ 2005).}, on the other, as a function of elemental
species.  {\boos} is included to contrast the abundance patterns of
the CEMP-no class with that of a ``normal'' extremely metal-poor RGB
star in the {\uf} dwarf galaxy {\boo}.  One sees very little
difference between {\boos} and the ``unmixed'' red giants of the
Galactic halo (see NYGW for a more detailed discussion).  In the
middle panel of the figure, the data for {\segs} have been connected
by a continuous line, which has been copied to the other panels to
facilitate comparison between {\segs} and the other objects.  The top
four panels show that CEMP-no stars comprise a somewhat heterogeneous
group.  First, [C/N] varies considerably among the Galactic halo
objects in the figure, from --1.4 in CS22949--037 to +0.2 in
CS22957--027, accompanied by a range in $^{12}$C/$^{13}$C from 4 to
10, respectively, both of which suggest that the CN(O) cycle(s) have
been more vigorous in the former than in the latter.  We note that
{\segs} has [C/N] = +1.5 and $^{12}$C/$^{13}$C $>$ 50, presumably
having experienced less extreme CN(O) processing.  Second, while
enhancement of Na, Mg, Al is frequently present in the CEMP-no class
(as first demonstrated for Mg and Al by Aoki et al.\ (2002)) and seen
here in {\segs}, no enhancement of [Mg/Fe] is evident in CS22957--027.

These differences notwithstanding, the relative abundances of {\segs},
taken as a whole, are consistent with its classification as a CEMP-no
star. What is the origin of this peculiar class of objects?  While
binarity has provided a most likely explanation for the CEMP-s
subclass (enhanced s-process neutron-capture elements; see
e.g. Lucatello et al.\ 2005), no such case has yet been generally
demonstrated for CEMP-no objects, and to the authors' knowledge only
CS22957--027 among the CEMP-no class exhibits direct evidence for
duplicity (Preston \& Sneden 2001).  A stronger argument against
binarity is that it offers no insight into the enhancements of Na, Mg,
and Al seen in a large fraction of the class.  A considerably more
successful, but somewhat {\it ad hoc}, explanation of the CEMP-no
phenomenon is the ``mixing-and-fallback'' model of Nomoto and
co-workers (see Nomoto et al.\ 2008, and references therein) which
postulates low energy supernova explosions at the earliest times, with
preferential expulsion of outer layers accompanied by fallback of
inner ones, resulting in the light elements C--Mg being relatively
more enhanced than the heavier ones.

That said, enrichment/mixing processes in star forming regions that
produce stars at [Fe/H] = --3.5 with [C/Fe] $>$ 1.5 and stars at
[Fe/H] = --2.0 with [C/Fe] = 0.0 (see Norris et al.~2010b) remain
obscure. {\seg} provides the first opportunity to quantify the
evolution of element ratios in a single system over the metallicity
range from extremely metal-poor with extreme carbon enhancement,
through extremely metal-poor with no carbon enhancement, to
metal-poor.

\subsection {Membership of {\seg}} 

The {\seg} system was discovered by Belokurov et al.\ (2007), who
identified it as ``an unusually extended globular cluster'', possibly
associated with the Sagittarius dSph.  Niederste-Ostholt et al.\
(2009) argued that determination of {\seg} membership is fraught with
difficulty by contamination of debris from the Sgr dSph.  Geha et al.\
(2009) and Simon et al.\ (2010), on the other hand, having obtained
accurate radial velocities for some 70 putative {\seg} members within
3.0{\rh} (half-light radii), inferred a total mass within {\rh} of 5.8
$\times$10$^{5}${\msun} together with M/L$_{V}$ = 3400{\msun}/{\lsun},
and concluded that it is a dark-matter-dominated dwarf galaxy.  They
also reported two members with [Fe/H] $<$ --3.3, together with stars
as metal-rich at [Fe/H] = --1.6, consistent with the abundance range
found by Norris et al.\ (2010b).

Here we center our attention on whether {\segs} is a member of the
{\seg} system.  First, we recall from Section 2.3 that its
heliocentric radial velocity of 204.3 $\pm$ 0.5 {\kms} is clearly
consistent with membership.  Second, it lies 16.9{\arcmin} from the
center of the system, corresponding to 3.8 half-light radii (r$_{h}$ =
4.4{\arcmin}, Martin et al.\ 2008).  If one adopts the surface
brightness distribution of Martin et al., the probability of
membership of {\segs} is $\sim$0.03 times that of an object lying at
half-light radius -- a moderately small likelihood.  Third, we recall
that in the range --4.0 $<$ [Fe/H] $<$ --3.0 CEMP-no red giants are
extremely rare in the Milky Way halo. For example, in the Beers,
Preston, \& Shectman (1992) survey for metal-poor stars there are some
14 red giants in this abundance range with $V$ $\la$ 14.5 and $B-V$
$>$ 0.60, only two of which (CS22949--037 and CS22957--027) are known
to be CEMP-no stars\footnote{We derive this from the high-resolution
abundance compilation of Frebel (2010,
https://www.cfa.harvard.edu/$\sim$afrebel/abundances/abund.html).},
over an area of 2300~deg$^{2}$.

\subsection{Fraction of CEMP stars as a Function of [Fe/H]}
 
It is well known that the fraction of CEMP stars in the Galactic halo
increases as [Fe/H] decreases, as summarized, for example, by Norris
et al.\ (2007, their Figure 1).  For stars having [Fe/H] $<$ --3.0 and
high-resolution, high-$S/N$, abundances, the fraction increases from
$\sim$ 0.1 at [Fe/H] = --3.0, to 1.0 below [Fe/H] $\sim$ --4.2.  It is
then of interest that while the number of stars in {\seg} having the
necessary observational data is small, the fraction of CEMP stars at
[Fe/H] $\sim$ --3.5 is 0.3.  We also note in passing that almost all
C-rich stars currently known with [Fe/H] $\la$ --3.3, and for which
[Ba/Fe] has been detected, belong to the CEMP-no subclass.  (Of
particular interest are the three stars with [Fe/H] $<$ --4.5, all of
which are C-rich: HE0557--4840 ([Fe/H] = --4.8) is CEMP-no (Norris et
al.\ 2007); HE0107--5240 ([Fe/H] = --5.4) has only an upper limit for
Ba of [Ba/Fe] $<$ +0.82 (Christlieb et al.\ 2002, 2004); and
HE1327--2326 ([Fe/H] = --5.6) has [Ba/Fe] $<$ +1.5, but [Sr/Fe] $\sim$
1.0 (Frebel et al.\ 2005; Aoki et al.\ 2006.)

Before concluding, we comment here on the CEMP classification.  As
clearly demonstrated by Aoki et al. (2007, their Figure 3), over the
range --4.0 $<$ [Fe/H] $<$ --1.0 the distribution of stars in the
([C/Fe], [Fe/H])--plane is bimodal, with a separation between C-rich
and C-normal subgroups at [C/Fe] $>$ --0.52$\times$[Fe/H] --0.23.  We
shall refer here to the C-rich subgroup as extremely carbon-rich,
noting that not all CEMP stars fall within this category -- at [Fe/H]
= --3.5, for example, an extremely C-rich star has [C/Fe] $>$ +1.6.
Insofar as we consider that Figure 3 of Aoki et al.\ (2007) provides a
strong constraint on the nature of carbon richness in metal-poor stars
we find this preferable to the somewhat arbitrary use of [C/Fe] $>$
+1.0 as a definition of C-rich stars.  With this background, we note
that some red giants with [Fe/H] $<$ --3.0 in other {\uf} dwarf
galaxies are known to contain moderate carbon enhancements: in the
range 0.5 $<$ [C/Fe] $<$ 0.8, Frebel et al.\ (2010b) find a C-rich
fraction of 2/3 for UMa II, while Norris et al.\ (2010b) report 1/3 for
{\boo}.  We do not regard these as extremely C-rich (i.e. above the
Aoki et al. (2007) threshold).

\subsection {Conclusion} 

What may one conclude?  First, {\segs} has a radial velocity
consistent with membership of the kinematic system defined by Simon et
al.\ (2010), which, taken together with the extreme rarity of CEMP-no
red giants, suggests it is not a member of the Galactic halo field.
{\segs} is thus associated with either the central concentration of
{\seg}, or with the more widely spread debris of Sgr advocated by
Niederste-Ostholt et al.\ (2009) to be a contaminant of the system.
Possibly inconsistent with membership of the central concentration is
the somewhat low probability that it is associated with the {\seg}
spatial distribution defined by Martin et al.\ (2008).  An
interpretation one might consider is that the Martin et al. radial
brightness profile is not an adequate description of the baryonic mass
distribution within the {\seg} system, which is instead better
described by the Niederste-Ostholt et al.\ (2009) photometric map,
with significant east-west extension.  That said, we consider it safe
to conclude that {\segs} originated in a dwarf galaxy system -- either
{\seg} or Sgr -- that contained extremely metal-poor material ([Fe/H]
$<$ --3.0).

{\segs} is the first extremely metal-poor, extremely C-rich star to be
associated with a dwarf galaxy satellite of the Milky Way, as opposed
to the general field of its halo.  Thus not only are there extremely
metal-poor, ``C-normal'' stars in the Galaxy's dwarf satellites with
abundance signatures similar to those of the Galaxy's halo, but also
objects with extreme carbon enhancements, [C/Fe] = +2.3, similar to
those found in the halo.  Such C-rich stars are extremely rare, and
membership (or past membership) of {\segs} in a dwarf galaxy now
located within the outer Milky Way shows that at extremely low
abundance, [Fe/H ] $<$ --3.0, chemical evolution proceeded similarly
in both the progenitors of the field halo and dwarf satellite systems,
consistent with the view that the latter played a role in the
formation of the Milky Way's outer halo.

\acknowledgements

 Studies at RSAA, ANU, of the most metal-poor stellar populations are
 supported by Australian Research Council grants DP0663562 and
 DP0984924, which J.E.N. and D.Y. gratefully acknowledge.
 R.F.G.W. acknowledges grants from the W.M. Keck Foundation and the
 Gordon \& Betty Moore Foundation, to establish a program of
 data-intensive science at the Johns Hopkins University, and NSF grant
 AST-0908326. A.F. is supported by a Clay Fellowship administered by
 the Smithsonian Astrophysical Observatory.\\

\noindent{\it Facilities:} {VLT:Kueyen(UVES)}

\clearpage                                                                                                    
                                                                                                              
\begin{deluxetable}{lllrrr}                                                                                   
\tablecolumns{6}                                                                                              
\tablewidth{0pt}                                                                                              
\tablecaption{\label{Tab:Linelist} EQUIVALENT WIDTHS, UPPER LIMITS, AND LINE-BY-LINE ABUNDANCES OF SEGUE 1--7}
\tablehead{                                                                                                   
  \colhead{}    & \colhead{$\lambda$} & \colhead{$\chi$} & \colhead{$\log gf$} &                              
  \colhead{$W_{\lambda}$} & \colhead{$\log\epsilon$} \\                                                       
 \colhead{Species} & \colhead{({\AA})}    & \colhead{(eV)}  & \colhead{(dex)}    &                            
  \colhead{(m{\AA})} & \colhead{(dex)} \\                                                                     
  \colhead{(1)} & \colhead{(2)}    & \colhead{(3)}  & \colhead{(4)}    &                                      
  \colhead{(5)} & \colhead{(6)} \\                                                                            
  }                                                                                                           
\startdata                                                                                                    
O  I  &  6300.304 &  0.00  &  $-$9.78  &   $<$10 &  $<$7.35 \\
Na I  &  5889.951 &  0.00  &   0.11  &   104 &   3.17 \\
Na I  &  5895.924 &  0.00  &  $-$0.19  &    92 &   3.20 \\
Mg I  &  3829.355 &  2.71  &  $-$0.21  &   148 &   5.07 \\
Mg I  &  3832.304 &  2.71  &   0.15  &   166 &   4.89 \\
Mg I  &  5172.684 &  2.71  &  $-$0.38  &   149 &   5.07 \\
Mg I  &  5183.604 &  2.72  &  $-$0.16  &   157 &   4.96 \\
Mg I  &  5528.405 &  4.34  &  $-$0.34  &    40 &   4.76 \\
Al I  &  3961.520 &  0.01  &  $-$0.34  &   102 &   3.08 \\
Ca I  &  4226.728 &  0.00  &   0.24  &   163 &   3.78 \\
Ca I  &  4318.652 &  1.90  &  $-$0.21  &    34 &   3.64 \\
Ca I  &  5265.556 &  2.52  &  $-$0.26  &    11 &   3.67 \\
Ca I  &  5588.749 &  2.52  &   0.21  &    24 &   3.62 \\
Ca I  &  5857.451 &  2.93  &   0.23  &    12 &   3.66 \\
Ca I  &  6102.723 &  1.88  &  $-$0.79  &    13 &   3.54 \\
Ca I  &  6122.217 &  1.89  &  $-$0.32  &    32 &   3.60 \\
Ca I  &  6162.173 &  1.90  &  $-$0.09  &    44 &   3.63 \\
Ca I  &  6439.075 &  2.52  &   0.47  &    34 &   3.56 \\
Ti I  &  3998.636 &  0.05  &  $-$0.06  &    30 &   1.91 \\
Ti II &  3759.296 &  0.61  &   0.27  &   113 &   1.96 \\
Ti II &  3761.323 &  0.57  &   0.17  &   108 &   1.91 \\
Ti II &  3913.468 &  1.12  &  $-$0.41  &    73 &   2.00 \\
Ti II &  4012.385 &  0.57  &  $-$1.75  &    41 &   1.80 \\
Ti II &  4443.794 &  1.08  &  $-$0.70  &    68 &   1.94 \\
Ti II &  4464.450 &  1.16  &  $-$1.81  &    31 &   2.27 \\
Ti II &  4468.507 &  1.13  &  $-$0.60  &    91 &   2.59 \\
Ti II &  4501.273 &  1.12  &  $-$0.76  &    55 &   1.70 \\
Ti II &  5188.680 &  1.58  &  $-$1.05  &    31 &   1.94 \\
Ti II &  5226.543 &  1.57  &  $-$1.23  &    26 &   1.99 \\
Ti II &  5336.771 &  1.58  &  $-$1.63  &    18 &   2.18 \\
Cr I  &  4254.332 &  0.00  &  $-$0.11  &    64 &   1.91 \\
Cr I  &  5206.038 &  0.94  &   0.02  &    27 &   1.87 \\
Cr I  &  5208.419 &  0.94  &   0.16  &    30 &   1.80 \\
Mn I  &  4033.062 &  0.00  &  $-$0.62  &    32 &   1.31\tablenotemark{a} \\
Fe I  &  3758.233 &  0.96  &  $-$0.03  &    96 &   3.61 \\
Fe I  &  3763.789 &  0.99  &  $-$0.24  &    91 &   3.72 \\
Fe I  &  3765.539 &  3.24  &   0.48  &    22 &   3.51 \\
Fe I  &  3787.880 &  1.01  &  $-$0.86  &    75 &   3.84 \\
Fe I  &  3815.840 &  1.48  &   0.24  &    97 &   3.90 \\
Fe I  &  3827.823 &  1.56  &   0.06  &    95 &   4.11 \\
Fe I  &  3865.523 &  1.01  &  $-$0.98  &    63 &   3.52 \\
Fe I  &  3899.707 &  0.09  &  $-$1.53  &    94 &   4.05 \\
Fe I  &  3922.912 &  0.05  &  $-$1.65  &    87 &   3.90 \\
Fe I  &  4005.242 &  1.56  &  $-$0.61  &    70 &   3.97 \\
Fe I  &  4045.812 &  1.48  &   0.28  &   102 &   3.90 \\
Fe I  &  4063.594 &  1.56  &   0.07  &    90 &   3.89 \\
Fe I  &  4071.738 &  1.61  &  $-$0.02  &    97 &   4.22 \\
Fe I  &  4132.058 &  1.61  &  $-$0.67  &    73 &   4.14 \\
Fe I  &  4143.868 &  1.56  &  $-$0.46  &    67 &   3.68 \\
Fe I  &  4233.603 &  2.48  &  $-$0.60  &    20 &   3.64 \\
Fe I  &  4260.474 &  2.40  &  $-$0.02  &    66 &   4.13 \\
Fe I  &  5171.596 &  1.49  &  $-$1.79  &    34 &   3.97 \\
Fe I  &  5191.455 &  3.04  &  $-$0.55  &    15 &   4.00 \\
Fe I  &  5194.942 &  1.56  &  $-$2.09  &    12 &   3.73 \\
Fe I  &  5216.274 &  1.61  &  $-$2.15  &    16 &   4.00 \\
Fe I  &  5232.940 &  2.94  &  $-$0.06  &    26 &   3.71 \\
Fe I  &  5266.555 &  3.00  &  $-$0.39  &    18 &   3.89 \\
Fe I  &  5269.537 &  0.86  &  $-$1.32  &    94 &   4.34 \\
Fe I  &  5324.179 &  3.21  &  $-$0.24  &    16 &   3.91 \\
Fe I  &  5328.039 &  0.92  &  $-$1.47  &    80 &   4.14 \\
Fe I  &  5328.532 &  1.56  &  $-$1.85  &    30 &   4.01 \\
Fe I  &  5371.490 &  0.96  &  $-$1.65  &    69 &   4.04 \\
Fe I  &  5397.128 &  0.92  &  $-$1.99  &    53 &   3.91 \\
Fe I  &  5405.775 &  0.99  &  $-$1.84  &    60 &   4.02 \\
Fe I  &  5429.697 &  0.96  &  $-$1.88  &    57 &   3.95 \\
Fe I  &  5434.524 &  1.01  &  $-$2.12  &    44 &   3.94 \\
Fe I  &  5446.917 &  0.99  &  $-$1.91  &    47 &   3.78 \\
Fe I  &  5455.609 &  1.01  &  $-$2.09  &    47 &   3.98 \\
Fe I  &  5497.516 &  1.01  &  $-$2.85  &    19 &   4.08 \\
Fe I  &  6137.692 &  2.59  &  $-$1.40  &    15 &   4.29 \\
Fe I  &  6393.601 &  2.43  &  $-$1.58  &    11 &   4.11 \\
Fe II &  4233.172 &  2.58  &  $-$1.90  &    25 &   3.79 \\
Fe II &  4923.927 &  2.89  &  $-$1.50\tablenotemark{b} &    40 &   4.04 \\
Co I  &  3845.461 &  0.92  &   0.01  &    61 &   2.12 \\
Co I  &  3995.302 &  0.92  &  $-$0.22  &    27 &   1.51 \\
Co I  &  4118.767 &  1.05  &  $-$0.49  &    25 &   1.90 \\
Co I  &  4121.311 &  0.92  &  $-$0.32  &    23 &   1.54 \\
Ni I  &  3858.292 &  0.42  &  $-$0.97  &    50 &   2.16 \\
Sr II &  4077.710 &  0.00  &   0.16  &    37 &  $-$2.18 \\
Sr II &  4215.520 &  0.00  &  $-$0.16  &    40 &  $-$1.81 \\
Ba II &  4934.076 &  0.00  &  $-$0.15  &   $<$10 & $<$$-$2.31 \\
Eu II &  4129.725 &  0.00  &   0.20    &   $<$10 & $<$$-$2.20 \\
\enddata                                                                                                      
\tablenotetext{a}{Value has been increased by 0.40 following Cayrel et al.\ (2004)}                           
\tablenotetext{b}{From VALD; Kupka et al.\ 1999 (\texttt{http://www.astro.uu.se/$\sim$vald/)}} 
\end{deluxetable}                                                                                             

\clearpage

\begin{deluxetable}{lrrrrr}                                                                                   
\tablecolumns{6}                                                                                              
\tablewidth{0pt}                                                                                              
\tablecaption{1D LTE Abundances of Segue 1--7}

\tablehead{                                                                                                   

\colhead{Species} & \colhead{N$_{\mbox{\scriptsize lines}}$} & \colhead{$\log\epsilon (\mbox{X})$} & \colhead{s.e.$_{\log\epsilon}$\tablenotemark{a}} & \colhead{[X/Fe]} & \colhead{$\sigma_{\rm tot}$} \\
\colhead{(1)} & \colhead{(2)}    & \colhead{(3)}  & \colhead{(4)}    &  \colhead{(5)} & \colhead{(6)}  

}                                                                                                           
\startdata 
  C(CH)       &  syn\tablenotemark{b} & $   7.17 $ & $   0.20 $ & $   2.30 $ & 0.30 \\
  N(NH)       &  syn\tablenotemark{b} & $   5.01 $ & $   0.30 $ & $   0.75 $ & 0.42 \\
  \ion{O}{1}  &    1 & $ < 7.35 $ & $    ... $ & $ < 2.21 $ &  ... \\
  \ion{Na}{1} &    2 & $   3.18 $ & $   0.04 $ & $   0.53 $ & 0.13 \\
  \ion{Mg}{1} &    5 & $   4.95 $ & $   0.07 $ & $   0.94 $ & 0.11 \\
  \ion{Al}{1} &    1 & $   3.08 $ & $    ... $ & $   0.23 $ & 0.19 \\
  \ion{Si}{1} &  syn\tablenotemark{b} & $   4.79 $ & $   0.20 $ & $   0.80 $ & 0.26 \\
  \ion{Ca}{1} &    9 & $   3.63 $ & $   0.04 $ & $   0.84 $ & 0.09 \\
  \ion{Ti}{1} &    1 & $   1.91 $ & $    ... $ & $   0.53 $ & 0.18 \\
  \ion{Ti}{2} &   11 & $   2.03 $ & $   0.08 $ & $   0.65 $ & 0.19 \\
  \ion{Cr}{1} &    3 & $   1.86 $ & $   0.05 $ & $  -0.26 $ & 0.10 \\
  \ion{Mn}{1} &    1 & $   1.31 $ & $    ... $ & $  -0.56 $ & 0.19 \\
  \ion{Fe}{1} &   37 & $   3.93 $ & $   0.03 $ & $  -3.52\tablenotemark{c} $ & 0.09 \\
  \ion{Fe}{2} &    2 & $   3.92 $ & $   0.16 $ & $  -0.01 $ & 0.25 \\
  \ion{Co}{1} &    4 & $   1.77 $ & $   0.15 $ & $   0.37 $ & 0.16 \\
  \ion{Ni}{1} &    1 & $   2.16 $ & $    ... $ & $  -0.55 $ & 0.18 \\
  \ion{Sr}{2} &    2 & $  -1.99 $ & $   0.23 $ & $  -1.39 $ & 0.27 \\
  \ion{Ba}{2} &    1 & $ <-2.31 $ & $    ... $ & $ <-0.96 $ &  ... \\
  \ion{Eu}{2} &    1 & $ <-2.20 $ & $    ... $ & $ < 0.80 $ &  ... \\

\enddata 
\tablenotetext{a}{Uncertainty of the fit in the case of spectrum
synthesis; standard error of the mean for species having at least two
line strength measurements}
\tablenotetext{b}{Determined using spectrum synthesis}
\tablenotetext{c}{The tabulated value is [FeI/H]}
\end{deluxetable} 

\clearpage
\begin{figure}[htbp]
\vspace{1cm}
\begin{center}
\includegraphics[width=8.0cm,angle=0]{f1.eps}

\caption{\label{Fig:Spectra} The upper panel presents spectra of
{\boos}, {\segs}, and {\cs} where for {\boos} the data were obtained
with VLT/UVES (Norris et al.\ 2010a) and for {\cs} with AAT/UCLES
(Norris et al.\ 1997).  All data have been smoothed with a Gaussian
having FWHM = 1.0~{\AA}.  Also shown in this panel are
{\teff}/{\logg}/[Fe/H]/[C/Fe] (from the cited and present works).  In
the lower panel the original unbroadened data are presented for
{\segs} and {\cs} over a smaller wavelength range.}

\end{center}
\end{figure}

\clearpage
\begin{figure}[htbp]
\vspace{0.cm}
\begin{center}
\includegraphics[width=6.0cm,angle=0]{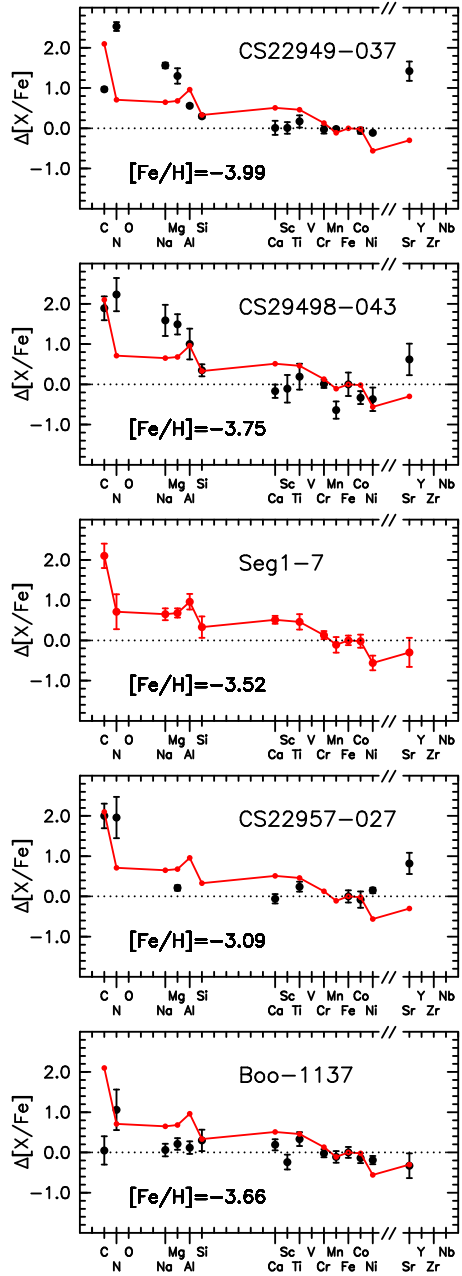}

\caption{\label{Fig:DXFe} Relative abundances differences,
$\Delta$[X/Fe], between C-rich stars (together with the C-normal
{\boos}) and the ``unmixed'' Galactic halo red giants of Spite et
al. (2005) as a function of species.  In the middle panel, the data
for {\segs} have been connected by a continuous line, which has been
copied to the other panels to facilitate comparison between it and
the other objects. Additional data sources are Aoki et al.\ (2004,
CS29498--043), Depagne et al.\ (2002) and Cayrel et al.\ (2004) (for
CS22949--037), Norris et al.\ (1997, CS22957--027), and Norris et al.\
(2010a, {\boos}).}

\end{center}
\end{figure}

\end{document}